\documentclass[11pt,twoside]{article}
\usepackage[utf8]{inputenc}
\usepackage[T1]{fontenc}
\usepackage[english]{babel}
\usepackage[paperheight=238mm,paperwidth=166mm,text={128mm,198mm},centering]{geometry}
\usepackage{amssymb,amsmath}
\usepackage{graphicx}
\usepackage{epstopdf}
\usepackage{scrextend}
\usepackage{setspace}
\usepackage{lipsum}
\usepackage{titlesec}
\usepackage[T1]{fontenc}
\usepackage{mathptmx}
\usepackage{fancyhdr}
\def \sitissue{%STATISTICS IN TRANSITION new series, Winter 2015
}
\def \sitshort{Kosiorowski D., Mielczarek D., Rydlewski J.P. and Snarska M.: Generalized...}

\titleformat{\section}
{\normalfont\bfseries\large}{\thesection.}{2mm}{}

\titleformat{\subsection}
{\normalfont\bfseries}{\thesubsection.}{2mm}{}

\begin{document}
\vspace*{-8mm}
\begin{small}
\noindent {\em \sitissue }\\
{\em Vol. XX, No. X, pp. xxx-xxx}
\end{small}

\begin{onehalfspace}
	\begin{center}
		{\Large \bf  GENERALIZED EXPONENTIAL SMOOTHING IN PREDICTION OF HIERARCHICAL TIME SERIES}
	\end{center}
\end{onehalfspace}

\vspace*{-4mm}

\begin{center}
	\begin{large}
	{\bf Daniel Kosiorowski}\footnote{ Department of Statistics, Cracow University of Economics, Krak\'ow, Poland. E-mail: daniel.kosiorowski@uek.krakow.pl},
	{\bf Dominik Mielczarek}\footnote{AGH University of Science and Technology, Faculty of Applied Mathematics, al. A. Mickiewicza 30, 30-059 Krakow, Poland. E-mail: dmielcza@wms.mat.agh.edu.pl}	
    {\bf Jerzy P. Rydlewski}\footnote{AGH University of Science and Technology, Faculty of Applied Mathematics, al. A. Mickiewicza 30, 30-059 Krakow, Poland. E-mail: ry@agh.edu.pl}
%% W afiliacjach AGH 1. ma byc najpierw AGH, a potem wydzial, 2. Krakow bez kreski nad o !!
    {\bf Ma\l gorzata Snarska}\footnote{Department of Financial Markets, Cracow University of Economics, Krak\'ow, Poland. E-mail: malgorzata.snarska@uek.krakow.pl}
    \end{large}
\end{center}

\begin{center}\begin{large} \textbf{ABSTRACT} \end{large} \end{center}

\begin{addmargin}[6mm]{6mm}
\begin{small}
\begin{singlespace}

Shang and Hyndman (2017) proposed a grouped functional time series forecasting approach as a combination of individual forecasts obtained using generalized least squares method. We modify their methodology using generalized exponential smoothing technique for the most disaggregated functional time series in order to obtain more robust predictor. We discuss some properties of our proposals basing on results obtained via simulation studies and analysis of real data related to a prediction of a demand for electricity in Australia in 2016.


\end{singlespace}
\end{small}
\end{addmargin}

\smallskip

\section{Introduction}

A problem of optimal reconciliation of forecasts of complex economic phenomena partitioned into certain groups and/or levels of hierarchy was considered in economic and econometric literature many times and is still present in a public economic debate (see Kohn (1982), Weale (1988), Hyndman et al. (2011)). National import/export quantities or Gross National Product balances are important examples here. Discrepancies between forecasts prepared at global level and obtained by aggregating regional forecasts or forecasts prepared according to certain hierarchy levels are usually thought to be caused by different methodologies or different precision of measurements used at different hierarchy levels or "prediction clusters".
The issue is also very important from a particular company point of view in a context of a product or a service lines management, consumers portfolio optimization and consumers segmentation. As an example of material product line management, let us take an equipment for running grouped in levels of hierarchy with respect to age, sex, competitive or re-creative usage and season designation. As examples of immaterial product sales optimization, let us take demand and supply of electricity within day and night optimized with respect to forecasted day and night demand,  customers grouped with respect to regions of living or residing and a degree of "consumer priority". As an example of a service management, let us take the Internet holiday booking service divided into sub-services with respect to certain wealth or "an inclination to adventures" criterion.     
\begin{figure}
\centering
\includegraphics[width=.95\linewidth]{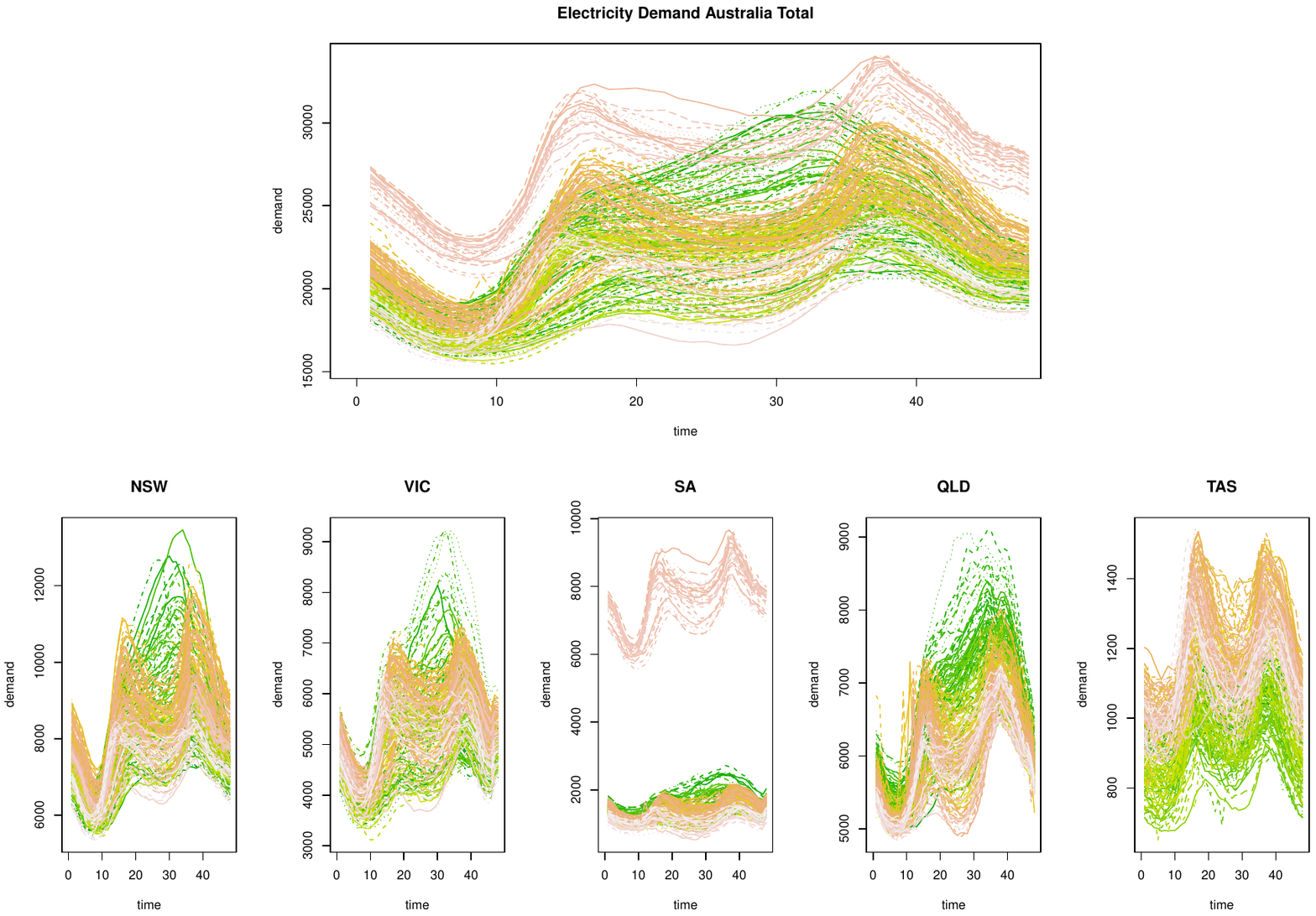}
\caption{Electricity demand in regions of Australia in 2016 -- hierarchical functional time series example}
\label{fig1}
\end{figure}
In recent years a very interesting statistical methodology named functional data analysis (FDA) for analyzing functional data has been developed (see Bosq (2000), Ramsay et al. (2009), Horvath and Kokoszka (2012), Krzy\'sko et al. (2013), Shang and Hyndman (2017)). For applications of the FDA in economics see Kosiorowski (2014), Kosiorowski (2016), Kosiorowski, Rydlewski and Snarska (2017a). Economic usefulness of outliers detection procedures in the FDA setup has been recently described and discussed in  Nagy et al. (2017), Kosiorowski, Rydlewski and Zawadzki (2018a), Kosiorowski, Mielczarek and Rydlewski (2018c).\\
In this context, it is worth stressing, that many others economic phenomena may effectively be described by means of functions or their series (i.a. utility curves, yield curves, development paths of companies or countries).\\ 

Following the above cited authors we consider a random curve $X=\{x(t), t\in [0,T]\},$ where $T$ is fixed, as a random element of the separable Hilbert space $L^2([0,T])$ with the inner product $<x,y>=\int x(t)y(t)dt$. The space is equipped with the Borel $\sigma-$algebra. Furthermore, in Bosq's (2000) monograph it is proved that probability distributions do exist for functional processes with values in Hilbert space. We denote this probability distribution by $\mathcal{F}$.
Functional time series (FTS) is a series of functions indexed by time (e.g., see Fig. 1, colors indicate time succession of sequence of the functional objects according to a base R  \textit{terrain.colors} color palette).
A hierarchical functional time series is a series of functions grouped at specified levels (household, town, region, whole country), (i.e., see Fig. 1). At each level a forecast can be made. A natural problem arises: how to use information obtained at different levels to obtain a reconciliated prediction for all levels?

The problem of hierarchical time series prediction is solved with various ways.
Bottom-up method relies on forecasting each of the disaggregated series at the lowest level of the hierarchy, and then using simple aggregation to obtain forecasts  at higher level of the hierarchy (see Kahn (1998)). Top-down method involves forecasting of aggregated series at the top level of the hierarchy, and then using  disaggregation to obtain forecasts at lower level of the hierarchy based on historical proportions.
Shang and Hyndman (2017), extending the method of Hyndman et al. (2011), considered grouped functional time series forecasting as an optimal combination of individual forecasts using generalized least squares regression with level forecasts treated as regressors.
In the context of hierarchical FTS prediction a general problem arises: which method of forecasting should be chosen (see Bosq (2000), Besse et al. (2000), Hyndman and Ullah (2007), Hyndman and Shang (2009), Aue et al. (2015), Kosiorowski, Mielczarek and Rydlewski (2017b, 2018b).
Shang and Hyndman (2017) proposed grouped functional time series forecasting approach as a combination of individual forecasts obtained by means of their smart predicting method, in which functional time series is reduced to family of one dimensional time series of principal component scores representing original functional series (see Kosiorowski, 2014). As a result of conducted simulation studies, we decided to modify their methodology. Instead of using principal component scores forecast methods, we decided to propose a certain functional generalization of exponential smoothing technique (see Hyndman et al. (2008) for a theoretical background of the exponential smoothing), i.e., we used moving local medians and moving local functional trimmed averages (Febrero-Bande and de la Fuente, 2012) for the most disaggregated series in order to obtain more robust predictor than Shang and Hyndman (2017).
Main aim of the paper is to modify Shang and Hyndman (2017) predictor so that it could cope with functional outliers and/or it would be elastic enough to adapt to changes in data generating mechanism.
The remainder of the paper is as follows: in the second section elements of depth concept for functional data are sketched and in the third section our proposals are introduced. Fourth section presents results of simulation as well as empirical studies. The paper ends with conclusions, references and short appendix containing R script showing how to calculate forecasts using our proposals with free \textit{DepthProc} and \textit{fda.usc} R package (see Kosiorowski and Zawadzki, 2018; Febrero-Bande and de la Fuente, 2012).

\section{Depths for functional data}
For obtaining robust hierarchical FTS predictor we focused our attention on the functional data depth concept (Nagy et al. (2016) and Nieto-Reyes and Battey (2016)). We have chosen, in our opinion the best depth for considered functional data, namely the corrected generalized band depth (cGBD, see L\'opez-Pintado and J\"{o}rnsten (2007)), but for computational reasons we restrict our considerations to the case of band consisting of two functions.\\
If $X_1$ and $X_2$ are independent functional random variables generated by the functional time series, and generating the observations (real functions in the $L^2([0,T])$ space), the cGBD of curve $x$ with respect to $\mathcal{F}$ is defined as
$$
cGBD(x|\mathcal{F})=\mathcal{F}\left(
G(x)\subset A^c(x;X_{1},X_{2})
\right)
$$
where $G(x)=\{(t,x(t)):t\in[0,T])\}$ is a graph of function $x$, and $a_{1,2}=\{t\in [0,T] : X_{2}(t)-X_{1}(t)\geq 0\}$ and
$$A^c(x;X_{1},X_{2})=
\{t\in a_{1,2} : X_{1}(t)\leq x(t)\leq X_{2}(t)\},  \textrm{ if } a_{1,2}\geq a_{2,1} $$
or
$$A^c(x;X_{1},X_{2})=\{t\in a_{2,1} : X_{2}(t)\leq x(t)\leq X_{1}(t)\},  \textrm{ if } a_{2,1}> a_{1,2}.
$$
Let now $\mathbf{X}^N=\{x_1,...,x_N\}$ be a sample of continuous curves defined on the compact interval $[0,T]$. Let $\lambda$ denote the Lebesgue measure and let $a(i_1,i_2)=\{t\in [0,T] : x_{i_2}(t)-x_{i_1}(t)\geq 0\}$, where $x_{i_1}$ and $x_{i_2}$ are band delimiting objects. Let $L_{i_1,i_2}=\frac{\lambda(a(i_1,i_2))}{\lambda([0,T])}$.
The empirical cGBD of a curve $x$ with respect to the sample $\mathbf{X}^N$, which estimates cGBD for curve $x$ in the considered functional space with respect to $\mathcal{F}$, is defined as (see L\'opez-Pintado and J\"{o}rnsten, 2007)
$$
cGBD(x|\mathbf{X}^N)=\frac {2}{N(N-1)}\sum_{1\leq i_1<i_2\leq N}\frac{\lambda(A^c(x;x_{i_1},x_{i_2}))}{\lambda([0,T])}
$$
where
$$A^c(x;x_{i_1},x_{i_2})=
\{t\in a(i_1,i_2) : x_{i_1}(t)\leq x(t)\leq x_{i_2}(t)\},  \textrm{ if } L_{i_1,i_2}\geq \frac 12 $$
or
$$A^c(x;x_{i_1},x_{i_2})=\{t\in a(i_2,i_1) : x_{i_2}(t)\leq x(t)\leq x_{i_1}(t)\},  \textrm{ if } L_{i_2,i_1}> \frac 12.
$$
Within this definition, the introduced earlier band depth (L\'opez-Pintado and Romo, 2009) is modified so that it only takes into account the proportion of the domain where the delimiting curves define a contiguous region which has non--zero width.
Further in our proposals we use \emph{the depth regions of order} $\alpha$ for considered cGBD, i.e.,
$R_{\alpha}(\mathcal{F})=\{x : cGBD(x,\mathcal{F})\geq\alpha\}.$ Note, that 
$\alpha$-central regions ${{R}_{\alpha }}(\mathcal{F})=\{x\in L^{2}([0,T]):D(x,\mathcal{F})\geq \alpha \}$ may be defined for any statistical depth function  $D(x,\mathcal{F})$, where $\mathcal{F}$ denotes a probability distribution (Zuo and Serfling, 2000).  Note also, that various robust and nonparametric descriptive characteristics, like scatter, skewness, kurtosis, may be expressed in terms of $\alpha-$ central regions. These regions are nested and inner regions contain less and less probability mass. Following Paindaveine and Van Bever (2013), when defining local depth it will be more appropriate to index the family $\{{{R}_{\alpha }}(\mathcal{F})\}$ by means of probability contents.
Consequently, for any $\beta \in (0,1]$ we define the smallest depth region with $\mathcal{F}$-probability equal or larger than $\beta$ as
$$R^{\beta}(\mathcal{F})=\bigcap_{\alpha\in A(\beta)}R_{\alpha}(\mathcal{F}), $$
where $A(\beta)=\{\alpha\geq0 : P(R_{\alpha}(\mathcal{F}))\geq\beta\}$.
The depth regions ${{R}_{\alpha }}(\mathcal{F})$ and ${{R}^{\beta }}(\mathcal{F})$ provide only the deepest point neighborhood.
In our considerations, we can replace probability distribution $\mathcal{F}=\mathcal{F}^{\mathbf{X}}$ by its symmetrized version for any function $x$, namely ${{\mathcal{F}}_x}=\frac{1}{2}{{\mathcal{F}}^{\mathbf{X}}}+\frac{1}{2}{{\mathcal{F}}^{2x-\mathbf{X}}}$.
For any depth function $D(\cdot ,\mathcal{F})$ the corresponding \emph{sample local depth function at the locality level $\beta \in (0,1]$} is $L{{D}^{\beta }}(x,{{\mathcal{F}}^{(N)}})=D(x,{{\mathcal{F}_x}^{\beta (N)}})$, where $\mathcal{F}_{x}^{\beta (N)}$ denotes the empirical probability distribution related to those functional observations that belong to $R_{x}^{\beta }({{\mathcal{F}}^{(N)}})$, where $\mathcal{F}^{(N)}$ denotes empirical probability distribution calculated from $X^N$. Thus $R_{x}^{\beta }({{\mathcal{F}}^{(N)}})$ is the smallest sample depth region that contains at least a proportion $\beta $ of the $2N$ random functions ${{x}_{1}},...,{{x}_{N}},2x-x_{1},...,2x-x_{N}$. Depth is always well defined -- it is an affine invariant from original depth. For $\beta =1$ we obtain global depth, while for $\beta \simeq 0$ we obtain extreme localization.
As in the population case, our sample local depth will require considering, for any $x\in L^{2}([0,T])$, the symmetrized distribution $\mathcal{F}_{x}^{(N)}$ which is empirical distribution associated with ${x_{1}},...,{x_{N}},2x-{x_{1}},...,2x-{x_{N}}$.
Sample properties of the (global) depths result from general findings presented in Zuo and Serfling (2000). Implementations of local versions of several depths including projection depth, Student, simplicial, $L^p$ depth, regression depth and modified band depth can be found in free R package \emph{DepthProc} (see Kosiorowski and Zawadzki, 2018). In order to choose the locality parameter $\beta$ we recommend using expert knowledge related to the number of components or regimes in the considered data. Sample properties of the local versions of depths result from general findings presented in Paindaveine and Van Bever (2013). For other concepts of local depths see, e.g., Sguera et al. (2016). 
\section{Our proposals}
We consider a sample of $N$ functions $\textbf{X}^N$=$\left\{ {{x}_{i}}(t): t\in [0,T], i=1,...,N\right\}$. Let
\\ $FD^{\beta}(y\textbf{|}\textbf{X}^N)$ denote sample functional depth of $y(t)$ with locality parameter $\beta$, e.g., the functional depth is equal to corrected generalized band depth: $FD=cGBD$. Sample $\beta$--local median is defined as
$$ME{{D}_{FD^\beta}}({\textbf{X}^{N}})=\underset{i=1,...,N}{\mathop{\arg \max }}\,FD^{\beta}({{x}_{i}}|{\textbf{X}^{N}}).$$
Assume now, that we observe a functional time series $x_{i}^{cluster}(t)$, $i=1,2,...$ for each considered level of hierarchy. We put our proposals forward.
\\
\vskip1mm
PROPOSAL 1: We independently apply on each considered level of hierarchy and in each cluster of that level a moving median predictor to obtain a forecast for the cluster, and then generalized exponential smoothing takes the following form
$${{\hat{x}}_{n+1}^{cluster}}(t)=ME{{D}_{FD^{\beta}}}({{W}_{n,k}^{cluster}}),$$
where ${{W}_{n,k}}$ denotes a moving window of length $k$ ending in a moment $n$, i.e., \\
${{W}_{n,k}^{cluster}}=\{{{x}_{n-k+1}^{cluster}}(t),...,{{x}_{n}^{cluster}}(t)\}$. \\
\vskip1mm
\hspace{2mm} Joint reconciliation of forecasts is conducted then and our reconciled predictor takes a form:
$$\hat{\mathbf{X}}_{n+1}(t)=\mathbf{F}(\hat{x}^{cluster_{1}}_{n+1}(t),...,\hat{x}^{cluster_{M}}_{n+1}(t)),$$
where $\mathbf{F}$ denotes reconciliation procedure. In the Proposal 1, $\mathbf{F}$ is a generalized least squares procedure originally proposed by Shang and Hyndman (2017) and $M$ is a number of moving median predictors.
\vskip1mm
PROPOSAL 2: 
Sample $\beta_{threshold}$--trimmed mean with locality parameter $\beta$ is defined as\\
$$ave(\beta_{threshold},\beta)(\textbf{X}^N)=ave(x_{i} : FD^{\beta}(x_i|\textbf{X}^N)>\beta_{threshold}),$$
where $ave$ denotes sample functional average, and $\beta_{threshold}$
is a a pre-specified trimming threshold.
In this setup a generalized exponential smoothing technique is applied independently on each considered level of hierarchy and in each cluster of that level as well. As a predictor for $(n+1)$\textit{th} moment we take
$${{\hat{x}}_{n+1}^{cluster}}(t)= z\cdot ave(\beta_{threshold}^{1},\beta_1)({{W}_{n_1,k_1}^{cluster}})+(1-z)\cdot ave(\beta_{threshold}^{1},\beta_1)({{W}_{n_2,k_2}^{cluster}}),$$
where ${{W}_{n,k}^{cluster}}$ denotes a moving window of length $k$ ending in a moment $n$ , i.e., ${{W}_{n,k}^{cluster}}=\{{{x}_{n-k+1}^{cluster}}(t),...,{{x}_{n}^{cluster}}(t)\}$, $z\in [0,1]$ is a forgetting parameter and $n_2<n_1$.
Thus we consider a closer past represented by ${{W}_{n_1}^{cluster}}$ and a further past represented by ${{W}_{n_2}^{cluster}}$.\\
Note that lengths of the moving windows $k,k_1,k_2$ used in Proposals 1 -- 2 relate to the analogous forgetting parameters $\alpha$ in the classical exponential smoothing. Additionally, we have at our disposal the resolution parameter $\beta$, at which we predict a phenomenon.
Such an approach allows us to accommodate expert knowledge and adjust forgetting and resolution parameters to researcher's requirements. For comparison purpose, notice that in original  Shang and Hyndman (2017) paper, the authors made predictions using functional regression based on constant in time functional principal components scores modelled by means of the well known one--dimensional time series models (see Hyndman and Shang, 2009). \\
In the next step we consider a whole hierarchy structure of the phenomenon.
 Assume that a hierarchical structure is described by fixed hierarchy levels, which are also divided into sub-levels, which are divided into sub-levels, and so on -- assume that we have $M$ clusters in the whole hierarchical structure.
Smart reconciliation of forecasts is conducted in this step and our reconciled predictor takes a form:
$$\hat{\mathbf{X}}_{n+1}(t)=\mathbf{F}(\hat{x}^{cluster_{1}}_{n+1}(t),...,\hat{x}^{cluster_{M}}_{n+1}(t)),$$
where $\mathbf{F}$ is a Generalized Least Squares Estimator (see Shang and Hyndman, 2017).
\\
Using Shang and Hyndman (2017) notation we can write our model in the 
form $$R_n=S_nb_n,$$ 
where $R_n$ is a vector of all series at all clusters, $b_n$ is a vector of the most disaggregated data and $S_n$ is a fixed matrix that shows a relation between them. In the considered example we have  $R_n=\left[R_{Australia}, \textit{   } R_{NSW},\textit{   } R_{QLD}, \textit{   }R_{SA},\textit{   } R_{TAS}, \textit{   }R_{VIC}\right]^T$, $b_n=\left[R_{NSW},\textit{   } R_{QLD}, \textit{   }R_{SA},\textit{   } R_{TAS}, \textit{   }R_{VIC}\right]^T$, where $T$ denotes a vector transpose. The matrix $S_n$ is an $6\times 5$ matrix, where the only non-zero elements are $S_{1i}=1$ for $i=1,2,...,5$ and $S_{jj-1}=1$ for $j=2,...,6$.
We propose to do the base forecast:
$$\hat{R}_{n+1}=S_{n+1}\beta_{n+1}+\epsilon_{n+1},$$ 
where $\hat{R}_{n+1}$ is a matrix of base forecast for all series at all levels,\\
$\beta_{n+1}=E[b_{n+1}|R_1,...,R_n]$ is the unknown mean of the forecast distribution of the most disaggregated series and $\epsilon_{n+1}$ is responsible for errors.
We propose to use a generalized least square method as in Shang and Hyndman (2017)
$$\hat{\beta}_{n+1}=\left(S^T_{n+1}W^{-1}S_{n+1}\right)^{-1}S^T_{n+1}W^{-1}\hat{R}_{n+1}$$
modified so that we use a robust estimator of the dispersion matrix $W$, i.e., instead of diagonal matrix, which contains forecast variances of each time series, we can use a robust measure of joint forecast dispersion taking into account dependency structure between the level forecasts. Notice that a dynamic updating of the dispersion matrix estimates should be considered in further studies.
Let us define our dispersion matrix:
$$W=diag\{v^{total},v^{cluster_{1}},...,v^{cluster_{M}}\}$$
where $$v^{cluster}=V \left\lbrace \int_0^T \left(x^{cluster}_{nk}(t)-\hat{x}^{cluster}_{n}(t)\right)^2 dt, k=1,2,...,K_{cluster}, n=1,2,...,N\right\rbrace,$$
$K_{cluster}$ is a number of observations in considered cluster in time $n$, $N$ is here a number of moments at which observations have been made 
and where $V$ is some chosen robust measure of dispersion. \emph{We propose to substitute $V=c \cdot MAD^2$ instead of Shang and Hyndman's (2017) variance}. If we consider a hierarchy as in the above Figure 1, our dispersion matrix takes a simple form:
$$W=diag\{v^{Australia},v^{NSW},v^{QLD},v^{SA},v^{TAS},v^{VIC}\}.$$ We propose to use $c \cdot MAD^2$ instead of variance or take into account a dependency structure between level series using the well known minimum covariance determinant (MCD) or recently proposed PCS robust matrix estimators of multivariate scatter (see Vakili and Schmitt (2014)).

\section{Properties of our proposals}

Thanks to kindness of prof. Han Lin Shang, who made his R script available for us, we calculated optimal combination of forecast predictor and we compared Shang and Hyndman (2017) proposal with our Proposals 1 and 2 and with independent moving functional mean (no reconciliation has been conducted for this predictor).

We generated samples of trajectories from one dimensional SV, GARCH, Wiener, Brownian bridge processes, functional FAR(1) processes tuned as in Didericksen et al. (2012), and various mixtures of them. In the cases of the considered mixtures, we treated one of the component as "good" and the rest as "bad" - outlying.
In the simulations we considered several locality parameters differing within the levels of hierarchy and several moving window lengths. We considered samples with and without functional magnitude as well as shape outliers (see Kosiorowski, Mielczarek, Rydlewski, 2018b, 2018c, and references therein). The outliers were defined with respect to the functional boxplot induced by the cGBD, i. e., we replaced 1\%, 5\%, 10\% of curves in the samples by means of arbitrary curves being outside a band determined by the functional boxplot whiskers, and compared medians and medians of absolute deviations from the medians (MAD) of integrated forecasts errors in these two situations. Fig. 2 presents simulated hierarchical functional time series consisted of three functional autoregression models of order 1 (i.e. FAR(1)) with gaussian kernels and sine--cosine errors design (see Didericksen et al. 2012). Fig. 3 presents corresponding level forecasts obtained by means of our Proposal 1 and local moving median calculated from 15-obs. windows and locality parameters equal to $0.45$. Fig. 4 presents simulated hierarchical time series consisted of three processes, each being mixtures of two one-dimensional stochastic volatility processes (SV). Fig. 5 presents corresponding level forecasts obtained by means of our Proposal 1 and local moving median calculated from 15-obs. window and locality parameters equal to $0.45$. 
In  Figures 2, 4, 5 and 6 colours indicate time sequence of the functional objects according to basic R package \textit{terrain.colors} colour palette.
We indicated an order of appearance of observations using colors palette starting from yellow and ending on blue. In the appendix we placed simple script depending on \textit{DepthProc} R package illustrating a general idea of the performed simulations.
\begin{figure}
\centering
\includegraphics[width=.95\linewidth]{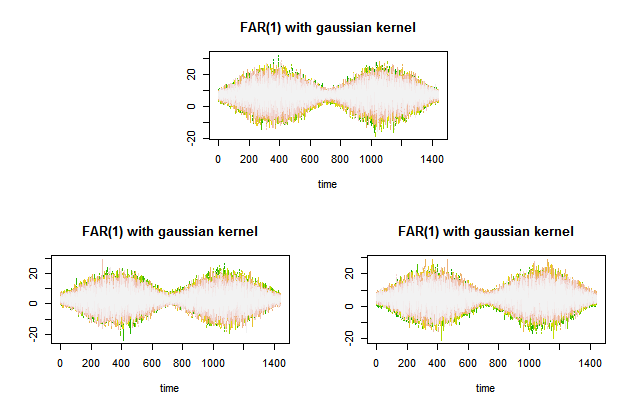}
\caption{Simulated HFTS consisted of FAR(1) processes}
\label{fig3}
\end{figure}
\begin{figure}
\centering
\includegraphics[width=.95\linewidth]{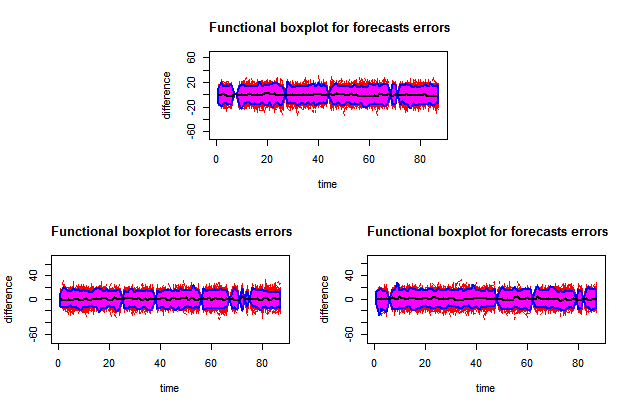}
\caption{HFTS prediction using our Proposal 1}
\label{fig4}
\end{figure}

\begin{figure}
\centering
\includegraphics[width=.95\linewidth]{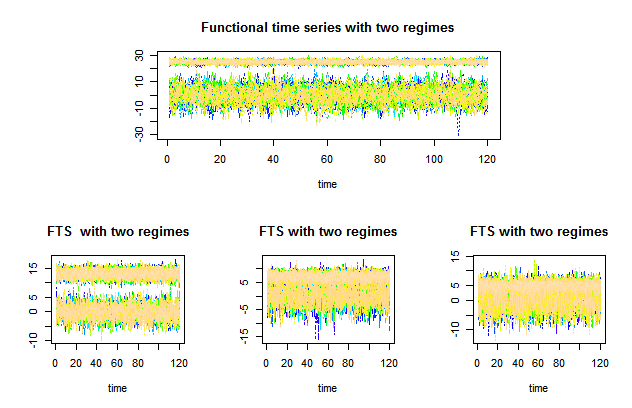}
\caption{Simulated HFTS consisted of two regime FTS processes}
\label{fig5}
\end{figure}
\begin{figure}
\centering
\includegraphics[width=.95\linewidth]{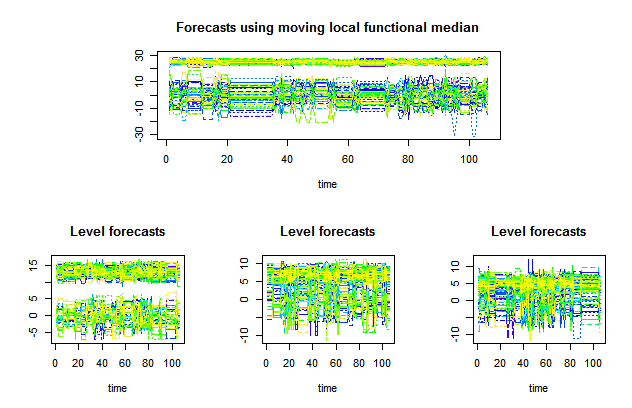}
\caption{HFTS prediction using our Proposal 1}
\label{fig6}
\end{figure}
In order to check the statistical properties of our proposals we considered empirical data set related to an electricity demand in the period from 1 to 31 January 2016 in Australia. The data come from five regions of Australia, denoted with the following symbols: NSW, QLD, SA, TAS, VIC. All the considered data was taken from Australian Energy Market Operator https://www.aemo.com.au/.
Fig. 6 presents 291 predicted electricity demand curves obtained by means of our Proposal 1 using moving local median to calculate level forecasts, window lengths $k=10$ observations and locality parameter equal to $\beta=0.2$. Fig. 7 presents boxplots for integrated prediction errors in each region and in the whole Australia. Fig. 8 presents functional boxplot for prediction of errors for whole Australia in 2016, where the predictions have been obtained using Proposal 1.
Fig. 9 presents functional boxplots for prediction of errors of electricity demand in NSW, QLD, VIC and TAS in 2016 obtained using Proposal 1. 
Note that functional median of prediction errors is close to $0$ for Australia and its regions (see Fig. 8 and 9). The boxplots in Fig. 7 show that a prediction error is small as well.
The volumes of central regions in Fig. 8 and 9 may be treated as predictor effectiveness measure.
We could therefore deduce that our predictor is median-unbiased (for more details on median-unbiasedness for functional data see Kosiorowski, Mielczarek and Rydlewski, 2017b, and references therein).

A performance of our proposals was compared with Shang and Hyndman (2017) proposal and with independent moving functional mean predictor (without the reconciliation of forecasts), in terms of a median of absolute deviation of integrated prediction errors in each region and in the whole Australia. Table 1 summarizes results of this comparison. In general, obtained results lead us to a conclusion, that our proposals seems to be more robust to functional outliers than Hyndman and Shang proposal. It is not surprising, as the authors made their forecasts basing on nonrobust generalized least squares method. Admittedly, Shang and Hyndman (2017) claimed that their proposal performed better in comparison to bottom-up approach basing on moving medians, but notice, that they considered Fraiman and Muniz global depths only. Moreover, thanks to the locality parameter adjusting, our proposals are more appropriate for detecting the change of regimes in the HFTS setup. In the cases of data sets without outliers, simple functional moving means, where the reconciliation procedure is not conducted, seems to outperform all other proposals. In this "clean data" situation, a performance of our both proposals and of Shang and Hyndman (2017) proposal is comparable. Our second proposal is more computationally demanding, however.
\begin{figure}
\centering
\includegraphics[width=.95\linewidth]{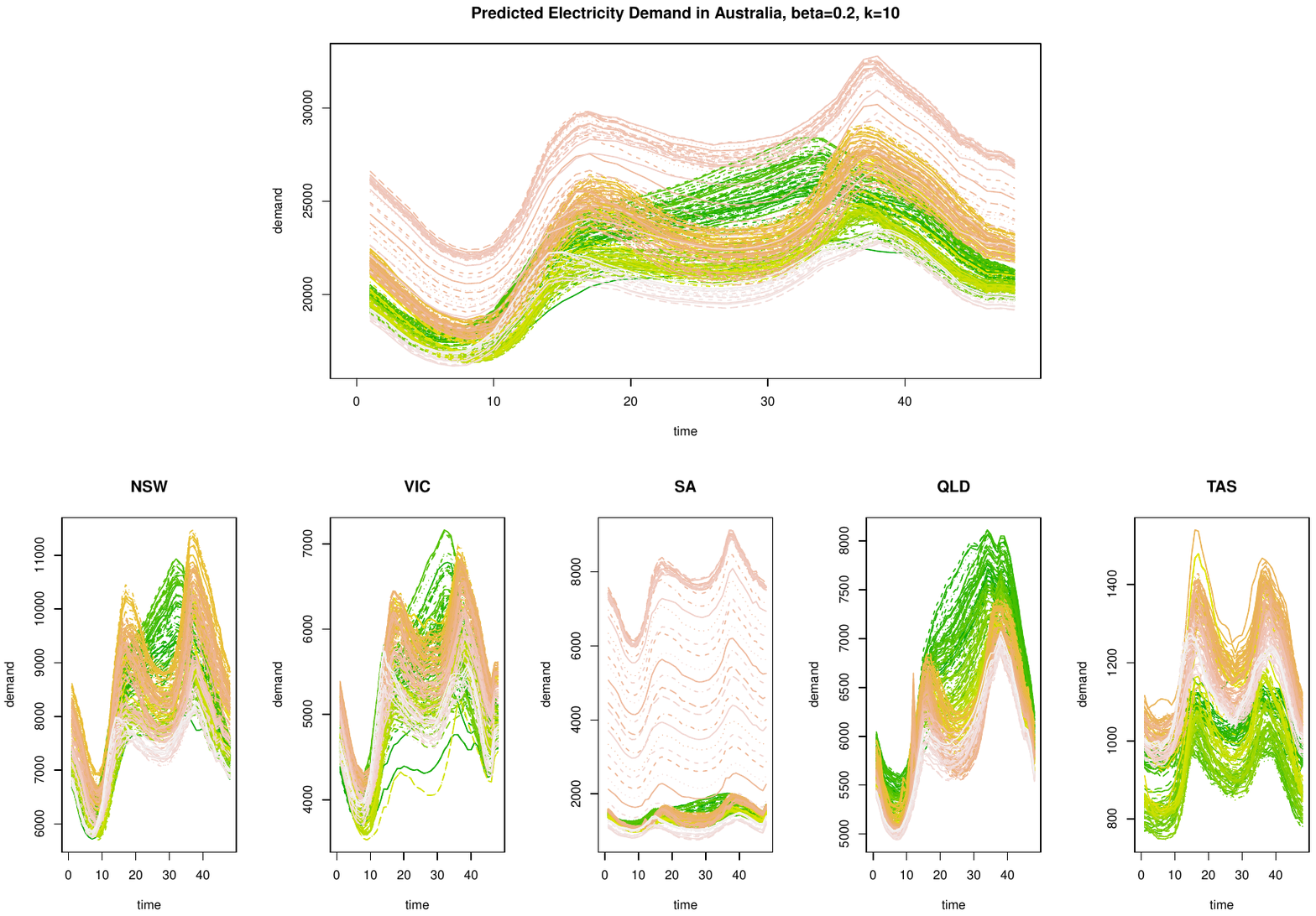}
\caption{Predicted electricity demand in Australia in 2016}
\label{fig7}
\end{figure}
\begin{figure}
\centering
\includegraphics[width=.95\linewidth]{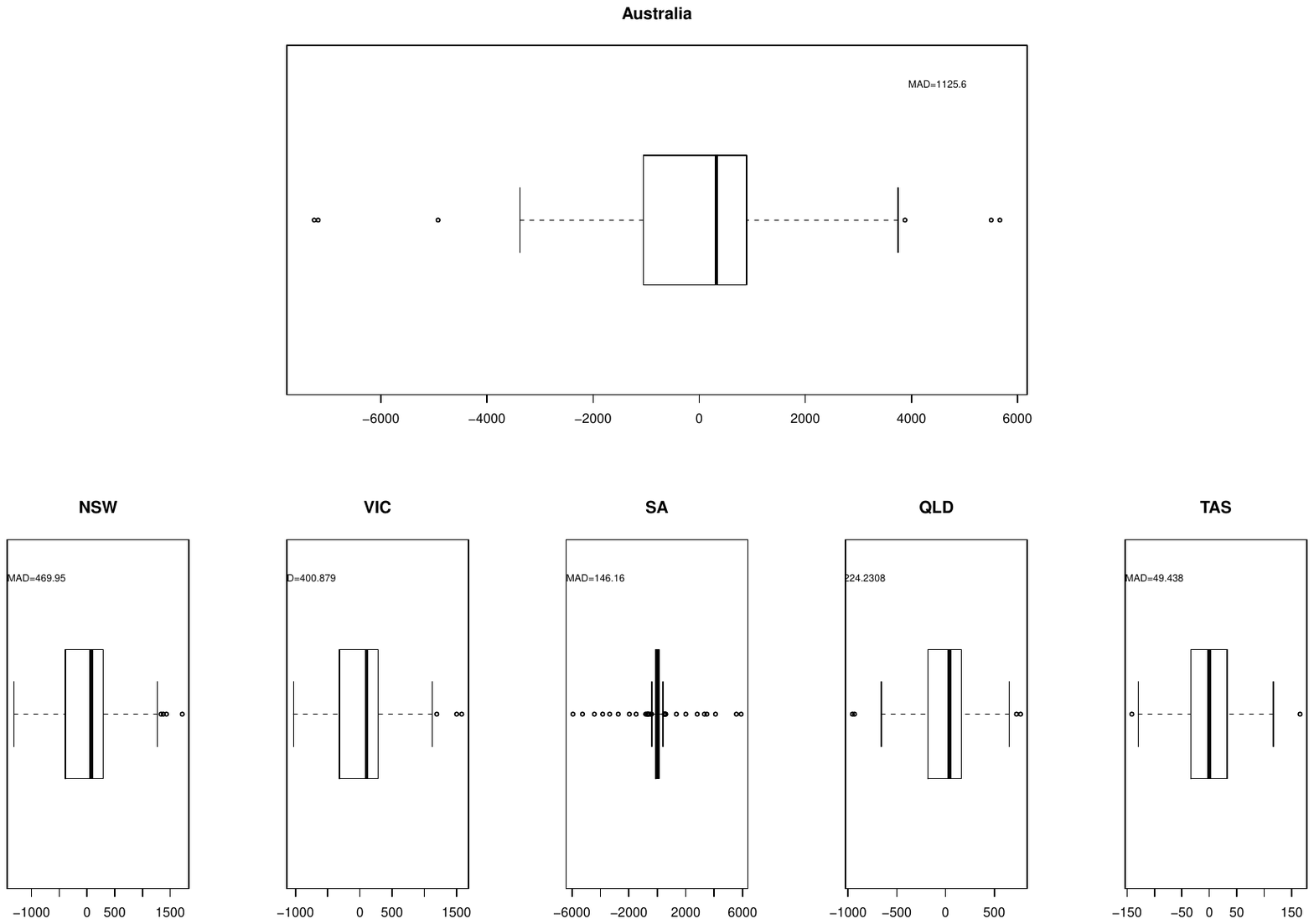}
\caption{Boxplots for integrated forecasts errors for electricity demand in Australia and its regions in 2016, predictions obtained using Proposal 1}
\label{fig8}
\end{figure}

\begin{figure}
\centering
\includegraphics[width=.88\linewidth]{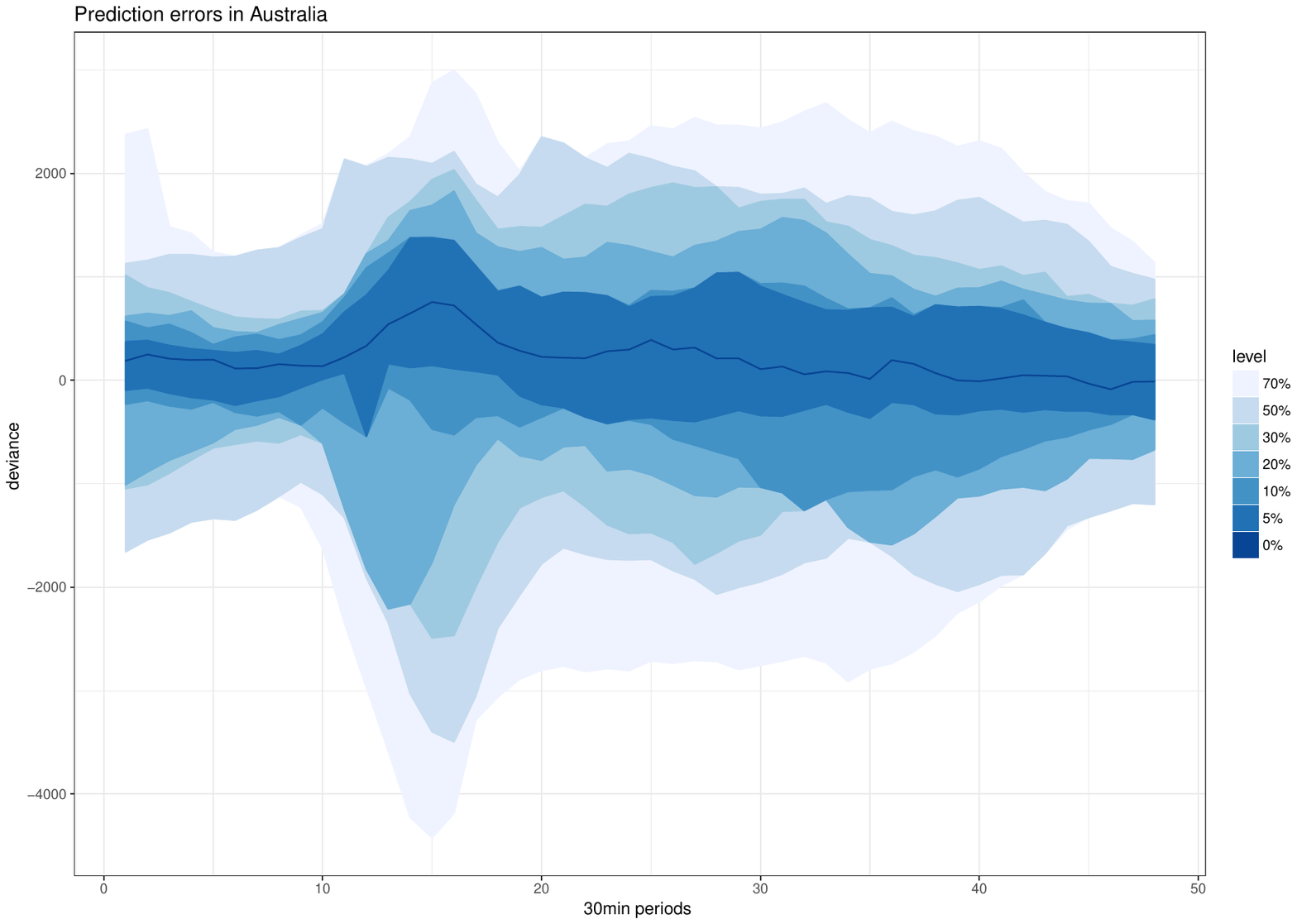}
\caption{Functional boxplot for prediction of errors for whole Australia in 2016, predictions obtained using Proposal 1}
\label{fig9}
\end{figure}
\begin{figure}
\centering
\includegraphics[width=.9\linewidth]{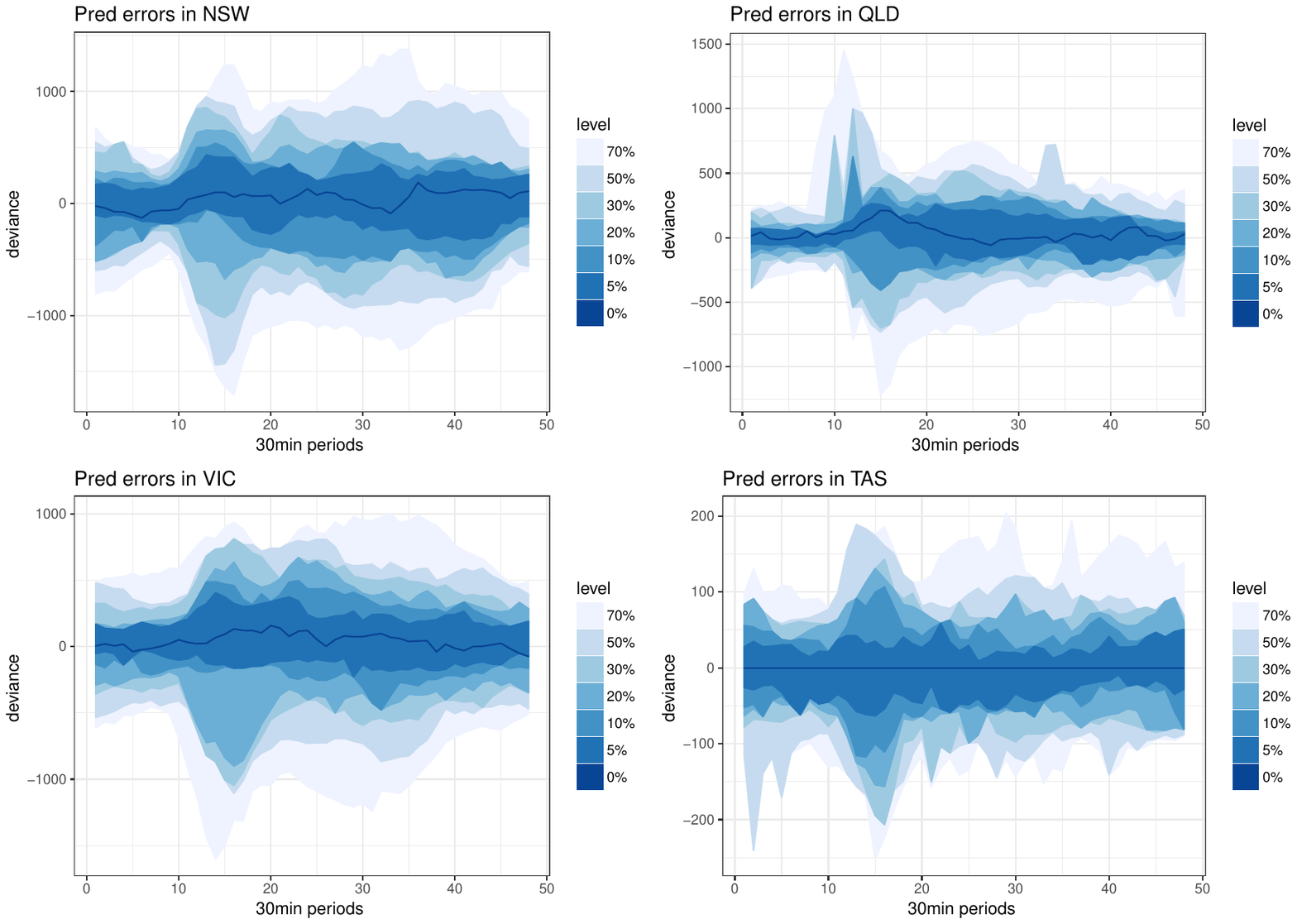}
\caption{Functional boxplots for prediction errors of electricity demand in NSW, QLD, VIC and TAS in 2016 obtained using Proposal 1}
\label{fig10}
\end{figure}
\begin{center}
\textbf{Table 1.} MAD of integrated forecasts errors     \\\smallskip
\begin{small}
\begin{tabular}{ccccccc}\hline
	$Predictor$ & $Austalia$ & $NSW$ & $VIC$ & $SA$ &$QLD$ & $TAS$ \\\hline
	Proposal 1 &  1126 & 470 & 401 & 146 & 224 & 49 \\
	Proposal 2 & 1311 & 628 & 452 & 147 & 181 & 52 \\
	H \& S Proposal & 1275 & 627 & 1004 & 176 & 230 & 51 \\
\hline	
\end{tabular}
\end{small}
\end{center}	
\subsection{Uncertainty evaluation}
Series of functional principal component scores are considered as
surrogates of original functional time series (see Aue et al. (2015), Hyndman and Shang (2009)). Several authors postulate using dynamic functional principal components approach in order to take into account the time changing dependency structure of described phenomenon (Aue et al., 2015). Notice, that such modification may drastically increase computational complexity of the HFTS procedure. In a context of uncertainty evaluation of our proposals, we suggest considering Vinod and L\'opez-de-Lacalle (2009) maximum entropy bootstrap for time series approach.
Bootstrap methods for FTS were studied among other by H\"ormann and Kokoszka (2012) and Shang (2018). Similarly, as in Shang and Hyndman (2017) we propose to use maximum entropy bootstrap methodology to obtain confidence regions and to conduct statistical inference. The \textit{meboot} and \textit{DepthProc} R packages give the appropriate computational support for that aims.

\section{Conclusions}
Hierarchical functional time series methodology opens new areas of statistical as well as economical research. E--economy provides a great deal of HFTS data. Our HFTS predictor proposal, which base on local moving functional median, performs surprisingly well in comparison to Shang and Hyndman (2017) proposal.
Lengths of the moving windows used in our proposals, relate to the forgetting parameters $\alpha$'s in the classical exponential smoothing. Moreover, we have at our disposal a "data resolution parameter" - $\beta$ at which we predict a phenomenon.
When using the locality parameter, we can take into account different sensitivity to details, e.g. number of different regimes of the considered phenomena. Further economic applications of the HFTS methodology may be found for example in Kosiorowski, Mielczarek, Rydlewski (2018b).
\section*{Acknowledgements}
JPR research has been partially supported by the AGH UST local grant no. 15.11.420.038.\\
DM and JPR research has been partially supported by the Faculty of Applied Mathematics AGH UST
statutory tasks within subsidy of Ministry of Science and Higher Education, grant no.
11.11.420.004.
\\
MS research was partially supported  by NSC Grant no. OPUS.2015.17.B.HS4.02708, DK research has been partially supported by the grant awarded to the Faculty of Management of CUE for preserving scientific resources for 2016, 2017 and 2018.

\noindent \begin{center}\begin{large}{\bf REFERENCES}\end{large}\end{center}

\everypar = {
\parindent=0pt
\hangindent=8mm
\hangafter=1
}\noindent

AUE, A., DUBABART NORINHO, D., H\"ORMANN, S. (2015). \textit{On the prediction of stationary functional time series}. \textit{Journal of the American Statistical Association}, 110(509), 378--392.

BESSE, P. C., CARDOT, H., STEPHENSON, D. B. (2000). \textit{Autoregressive forecasting of some functional climatic variations}. \textit{Scandinavian Journal of Statistics}, 27(4), 673--687.

BOSQ, D. (2000). \textit{Linear processes in function spaces}. Springer.

DIDERICKSEN, D., KOKOSZKA, P., ZHANG, X. (2012). \textit{Empirical properties of forecasts with the functional autoregressive model}. \textit{Computational Statistics}, 27(2), 285--298.

FEBRERO-BANDE, M. O., DE LA FUENTE, M. (2012). \textit{Statistical computing in functional data Analysis: the R package fda.usc}, \textit{Journal of Statistical Software}, 51(4), 1--28.

HORVATH, L., KOKOSZKA, P. (2012). \textit{Inference for functional data with applications}, Springer-Verlag.

H\"ORMANN S., KOKOSZKA, P. (2012). \textit{Functional Time Series}, in \textit{Handbook
of Statistics: Time Series Analysis -- Methods and Applications}, 30, 157–186.

HYNDMAN, R. J., AHMED R. A., ATHANASOPOULOS, G.,  SHANG, H. L. (2011). \textit{Optimal combination forecasts for hierarchical time series}, \textit{Computational Statistics \& Data Analysis}, 55(9), 2579 --2589.

HYNDMAN, R.J., KOEHLER, A.B., ORD, J.K., SNYDER, R.D. (2008). \textit{Forecasting with exponential smoothing -- the state space approach}, Springer-Verlag.

HYNDMAN, R. J., SHANG, H., L. (2009). \textit{Forecasting functional time series}, \textit{Journal of the Korean Statistical Society}, 38(3), 199 --221.

HYNDMAN, R.J., ULLAH, M. (2007). \textit{Robust forecasting of mortality and fertility rates: A functional data approach}, \textit{Computational Statistics \& Data Analysis}, 51(10), 4942 -- 4956.

HYNDMAN, R.J., KOEHLER, A.B., SNYDER, R.D., GROSE, S. (2002). A state space framework for automatic forecasting using exponential smoothing methods. \textit{International Journal of Forecasting}, 18(3), 439--454.

KAHN, K.B. (1998). \textit{Revisiting top-down versus bottom-up forecasting}, \textit{The Journal of Business Forecasting Methods \& Systems}, 17(2), 14--19.

KOHN, R. (1982). \textit{When is an aggregate of a time series efficiently forecast by its past}, \textit{Journal of Econometrics}, 18(3), 337 -- 349.

KOSIOROWSKI, D., ZAWADZKI, Z. (2018). \textit{DepthProc: An R package for robust exploration of multidimensional economic phenomena},
\emph{arXiv: 1408.4542}.

KOSIOROWSKI, D. (2014). \textit{Functional regression in short term prediction of economic time series}, \textit{Statistics in Transition}, 15(4), 611 --626.

KOSIOROWSKI, D. (2016). \textit{Dilemmas of robust analysis of economic data streams}, \textit{Journal of Mathematical Sciences} (Springer), 218(2), 167--181,  2016.

KOSIOROWSKI, D., RYDLEWSKI, J.P., SNARSKA M. (2017a). \textit{Detecting a structural change in functional
time series using local Wilcoxon statistic}, \textit{Statistical Papers}, pp. 1–22, URL http://dx.doi.org/10.1007/
s00362-017-0891-y.

KOSIOROWSKI, D., MIELCZAREK, D., RYDLEWSKI, J.P. (2017b). \textit{Double functional
median in robust prediction of hierarchical functional time series -- an
application to forecasting of the Internet service users behaviour}, available at:
arXiv:1710.02669v1.

KOSIOROWSKI, D., RYDLEWSKI, J.P., ZAWADZKI Z. (2018a). \textit{Functional outliers
detection by the example of air quality monitoring}, \textit{Statistical Review} (in Polish,
forthcoming).

KOSIOROWSKI, D., MIELCZAREK, D., RYDLEWSKI, J.P. (2018b). \textit{Forecasting of a Hierarchical
Functional Time Series on Example of Macromodel for the Day and Night Air Pollution in
Silesia Region - A Critical Overview}, 
\textit{Central European Journal of Economic Modelling and Econometrics}, 10(1), 26-46.

KOSIOROWSKI, D., MIELCZAREK, D., RYDLEWSKI, J.P. (2018c). \textit{Outliers in Functional Time Series -- Challenges for Theory and
Applications of Robust Statistics}, In M. Papież \& S. Śmiech (eds.), \textit{The $12^{th}$ Professor Aleksander Zelias International Conference on Modelling and Forecasting of Socio-Economic Phenomena. Conference  Proceedings}. Cracow: Foundation  of  the  Cracow 
University of Economics, to appear.

KRZY\'SKO, M., DERĘGOWSKI, K., GÓRECKI, T., WOŁYŃSKI,  W. (2013). \textit{Kernel and functional principal component analysis}, \textit{Multivariate Statistical Analysis 2013 Conference}, plenary lecture.

L\'OPEZ-PINTADO, S., ROMO, J. (2009).
 \textit{On the concept of depth for functional data}, \textit{Journal of the American Statistical Association}, 104, 718-734.

L\'OPEZ-PINTADO, S., J\"{O}RNSTEN, R. (2007). \textit{Functional analysis via extensions of the band depth}, IMS Lecture Notes--Monograph Series Complex Datasets and Inverse Problems: Tomography, Networks and Beyond, Vol. 54, 103--120, Institute of Mathematical Statistics.

NAGY, S., GIJBELS, I., OMELKA, M., HLUBINKA, D. (2016). \textit{Integrated depth for functional data: Statistical properties and consistency}, \textit{ESIAM Probability and Statistics}, 20, 95-130.

NAGY, S. GIJBELS, I., HLUBINKA, D. (2017). \textit{Depth-Based
Recognition of Shape Outlying Functions}, \textit{Journal of Computational and Graphical Statistics}, DOI:
10.1080/10618600.2017.1336445

NIETO-REYES, A., BATTEY, H. (2016). \textit{A Topologically valid definition of depth for functional data}, \textit{Statistical Science} 31(1), 61--79.

PAINDAVEINE, D., G. VAN BEVER, G. (2013). \textit{From depth to local depth: A Focus on centrality}, \textit{Journal of the American Statistical Asssociation}, Vol. 108, No. 503, Theory and Methods, 1105--1119.

RAMSAY, J.O., G. HOOKER, G., GRAVES, S. (2009). Functional data analysis with R and Matlab, Springer-Verlag.

SGUERA, C., GALEANO, P., LILLO, R.E. (2016). \textit{Global and local functional depths}, arXiv 1607.05042v1.

SHANG, H., L., HYNDMAN, R.J. (2017).
\textit{Grouped functional time series forecasting: an application to age-specific mortality rates}, \textit{Journal of Computational and Graphical Statistics}, 26(2), 330–343.

SHANG, H., L. (2018). \textit{Bootstrap methods for stationary functional time series}, \textit{Statistics and Computing}, 28(1), 1-10.

WEALE, M. (1988). \textit{The reconciliation of values, volumes and prices in the national accounts}, \textit{Journal of the Royal Statistical Society A}, 151(1), 211--221.

 VAKILI, K., SCHMITT, E. (2014). \textit{Finding multivariate outliers with FastPCS}. \textit{Computational Statistics \& Data Analysis}, 69, 54--66.
 
VINOD, H.D., L\'OPEZ-DE-LACALLE, J. L. (2009). \textit{Maximum entropy bootstrap for time series: the meboot R package}, \textit{Journal of Statistical Software}, 29(5).

ZUO, Y., SERFLING, R. (2000). \textit{Structural properties and convergence results for contours of sample statistical depth functions}, \textit{Annals of Statistics}, 28(2), 483--499.

Australian Energy Market Operator https://www.aemo.com.au/
\section*{Appendix}
\small
\#Simple R script, example showing how to calculate base forecasts for three hierarchy levels
\#using moving functional median implemented within the DepthProc R package.

require(DepthProc)\\
require(fda.usc)\\
require(RColorBrewer)\\
require(zoo)\\
\vskip1mm
wrapMBD = function(x)$\{$
  depthMedian(x,
  $depth\_params$ = list(method="Local", beta=0.45, depth\_params1 = list(method = "MBD")))
\}
\vskip1mm
\#Simple stochastic volatility 1D process simulator\\
SV <- function(n, gamma, fi, sigma, delta) \{

epsilon <- rnorm(n)

eta <- rnorm(2*n, 0, delta)

h <- c()

h[1] <- rnorm(1)

for (t in 2:(2*n)) \{

h[t] <- exp(gamma+fi*(h[t-1]-gamma)+sigma*eta[t])
\}

Z <- sqrt(tail(h,n)) * epsilon\\
return(Z)\}

example <- SV(100, 0, 0.2, 0.5, 0.1)

plot(ts(example))

\#functional time series simulator

m.data1<-function(n,a,b)
\{

M<-matrix(nrow=n,ncol=120)

for (i in 1:n) M[i,]<- a*SV(120,0,0.3,0.5,0.1)+b

M
\}

m.data.out1<-function(eps,m,n,a,b,c,d)\{

H<-rbind(m.data1(m,a,b),m.data1(n,c,d))

ind=sample((m+n),eps)

H1=H[ind,]

H1
\}

m <- matrix(c(1, 0, 1,  3, 2, 3, 2, 0), nrow = 2, ncol = 4)
m[2,]=c(2,2,3,3)
m[1,]=c(0,1,1,0)

\#below three functional time series

M2A= m.data.out1(150,3000,7000,5,0,1,25)

M2B= m.data.out1(150,3000,7000,2,0,1,15)

M2C= m.data.out1(150,3000,7000,3,0,1,10)

matplot(t(M2A),type="l",col=topo.colors(151),xlab="time",
main="Functional time series with two regimes")

matplot(t(M2B),type="l",col=topo.colors(151),xlab="time",
main="FTS  with two regimes")

matplot(t(M2C),type="l",col=topo.colors(151), xlab="time",
main="FTS with two regimes")

\#below moving local medians applied to the above series,
window lengths = 15 obs.,

\#locality parameters betas = 0.45

result4A = rollapply(t(M2A),width = 15, wrapMBD,
by.column = FALSE)

result4B = rollapply(t(M2B),width = 15,wrapMBD,
by.column = FALSE)

result4C = rollapply(t(M2C),width = 15,  wrapMBD,
by.column = FALSE)

matplot(result4A,type="l",col=topo.colors(87),
xlab="time",main="local 15-obs moving functional median, beta=0.45")

matplot(result4B,type="l",col=topo.colors(87), xlab="time",main="local 15-obs moving functional median, beta=0.45")

matplot(result4C,type="l",col=topo.colors(87),
xlab="time",main="local 15-obs moving functional median,\\
beta=0.45")\\
\vskip1mm
\#basic function for calculating $\beta\_treshold-$ trimmed $\beta-$ local MBD functional mean\\
\vskip1mm
beta\_tresh\_mean<-function(x,beta\_tresh,beta){\\
depths= depth(x, depth\_params = list(method="Local",
beta=beta, depth\_params1 = list(method = "MBD")))\\
ind=which(depths>beta\_tresh)\\
wyn= func.mean(x[ind,])\\
wyn\$data}

\end{document}